# Characterizing bright δ Scuti pulsators using *TESS* light curves

Prasad Mani,[1★] Timothy R. Bedding ,[1] Mara Bernizzoni,[1] Simon J. Murphy[2] and Daniel Hey[3]

[1] *Sydney Institute for Astronomy, School of Physics, University of Sydney, Sydney, NSW 2006, Australia*
[2] *Centre for Astrophysics, University of Southern Queensland, Toowoomba, QLD 4350, Australia*
[3] *Institute for Astronomy, University of Hawaii, Honolulu, HI 96822, USA*

## ABSTRACT

We present a study of bright, young δ Scuti stars near the zero-age main sequence using *TESS* light curves and *Gaia* DR3 data. From a sample of 2041 stars with $G < 7$ and $G_{BP} - G_{RP}$ colour in the range 0–0.6, we identified 444 δ Scuti pulsators. We measured a pulsator fraction of ∼70 per cent in the middle of the instability strip, tapering off towards the edges. A period–luminosity diagram reveals a concentration along the fundamental mode and overtone ridges. We addressed sample completeness and identified low-frequency pulsators possibly exhibiting mixed modes. Cross-matching with nearby young associations showed that 63 δ Scuti stars from our sample are association members.

## 1 INTRODUCTION

Asteroseismology, the science of inferring fundamental properties of stars from their pulsations (Aerts, Christensen-Dalsgaard & Kurtz 2010), has driven significant advancements in our understanding of stellar structure in many classes of stars (García & Ballot 2019; Aerts 2021; Kurtz 2022). It has enabled tighter constraints on temperature (Huber et al. 2012), age (e.g. Lebreton & Goupil 2014; Fritzewski et al. 2024), rotation (Aerts 2015), and metallicity (Kurtz 2022).

The category of intermediate-mass pulsating stars (1.2–2.5 $M_{\odot}$) called δ Scuti stars are found along the main sequence at the bottom of the instability strip in the Hertzsprung-Russel (H–R) diagram (Guzik 2021). They pulsate in low-radial-order pressure modes, where mode excitation is thought to be driven by the $\kappa$ mechanism (Cox 1980; Guzik, Fontes & Fryer 2018) in the He II partial ionization layers, with some contribution to the driving from turbulent pressure (Antoci et al. 2014) and the edge-bump mechanism (Stellingwerf 1979; Murphy et al. 2020b). Although δ Scuti oscillations are coherent with long-lived, high Signal to Noise (S/N) peaks in the amplitude spectra, the excited modes usually do not show any discernible pattern. Hence, identifying modes (assigning correct quantum numbers to various peaks) is difficult, inhibiting efforts to perform asteroseismology. Nevertheless, attempts have been made to understand the place of δ Scuti stars in the field of stellar pulsations (e.g. Balona, Daszyńska-Daszkiewicz & Pamyatnykh 2015; Bowman et al. 2016; Moya et al. 2017; Sánchez Arias, Córsico & Althaus 2017; Bowman & Kurtz 2018; Streamer et al. 2018). One important question which has been asked and studied before (Murphy et al. 2019; Gootkin et al. 2024) is why pulsations are only seen in a fraction of stars in the region of the H–R diagram where δ Scuti stars are found.

The *Transiting Exoplanet Survey Satellite* (*TESS*) (Ricker et al. 2015) is proving a powerful tool for studying δ Scuti stars. It was launched in 2018 and observes the sky in 27-d sectors at several different cadences[1] (20, 120, 200, 600, 1800 s). Since its pixel size is relatively large (≃ 21 arcsec[2]), *TESS* covers a large field on the sky in each sector. Supplementing the *Kepler*/K2 era population studies of δ Scuti stars (see reviews by Bowman & Kurtz 2018; Guzik 2021), the wide-field photometry of *TESS* has facilitated ensemble study of δ Scuti stars (for instance, Antoci et al. 2019; Balona & Ozuyar 2020; Barceló Forteza et al. 2020; Bedding et al. 2020; Murphy et al. 2020a; Skarka et al. 2022; Gootkin et al. 2024; Olmschenk et al. 2024; Skarka & Henzl 2024). Research using *TESS* has included searching for δ Scuti pulsations in pre-main-sequence stars (Murphy et al. 2021; Steindl, Zwintz & Bowman 2021), constraining ages of young open cluster $\alpha$ Per using δ Scuti stars (Pamos Ortega et al. 2022), characterizing δ Scuti stars in the Pleiades open cluster (Bedding et al. 2023), studying δ Scuti stars in the newly discovered Cep-Her association (Murphy et al. 2024), and others (e.g. Rodríguez-Martín et al. 2020; Hey et al. 2021; Murphy et al. 2022; Pamos Ortega et al. 2023; Dholakia et al. 2025), with the more recent studies being facilitated greatly by *Gaia* astrometry and photometry (Riello et al. 2021).

This is the second in a series of papers examining δ Scuti stars with *TESS* over the whole sky. Rather than finding as many pulsators as possible, our approach is to restrict each study to a well-defined sample. The first paper (Read et al. 2024) examined ∼1700 stars in a narrow colour range in the centre of the δ Scuti instability strip ($0.29 < G_{BP} - G_{RP} < 0.31$), restricted to those within 500 pc of the Sun (mostly brighter than $G = 11$). The fraction of bright stars ($G < 8$) showing δ Scuti pulsations was about 70 per cent, and dropped to about 45 per cent for fainter stars. An important conclusion was that a single sector of *TESS* data only detects the lowest-amplitude $\delta$ Scuti pulsations (around 50 ppm) in stars down to about $G = 9$.

★ E-mail: prasadmani94@gmail.com

[1] https://tess.mit.edu/science/
[2] https://heasarc.gsfc.nasa.gov/docs/tess/the-tess-space-telescope.html

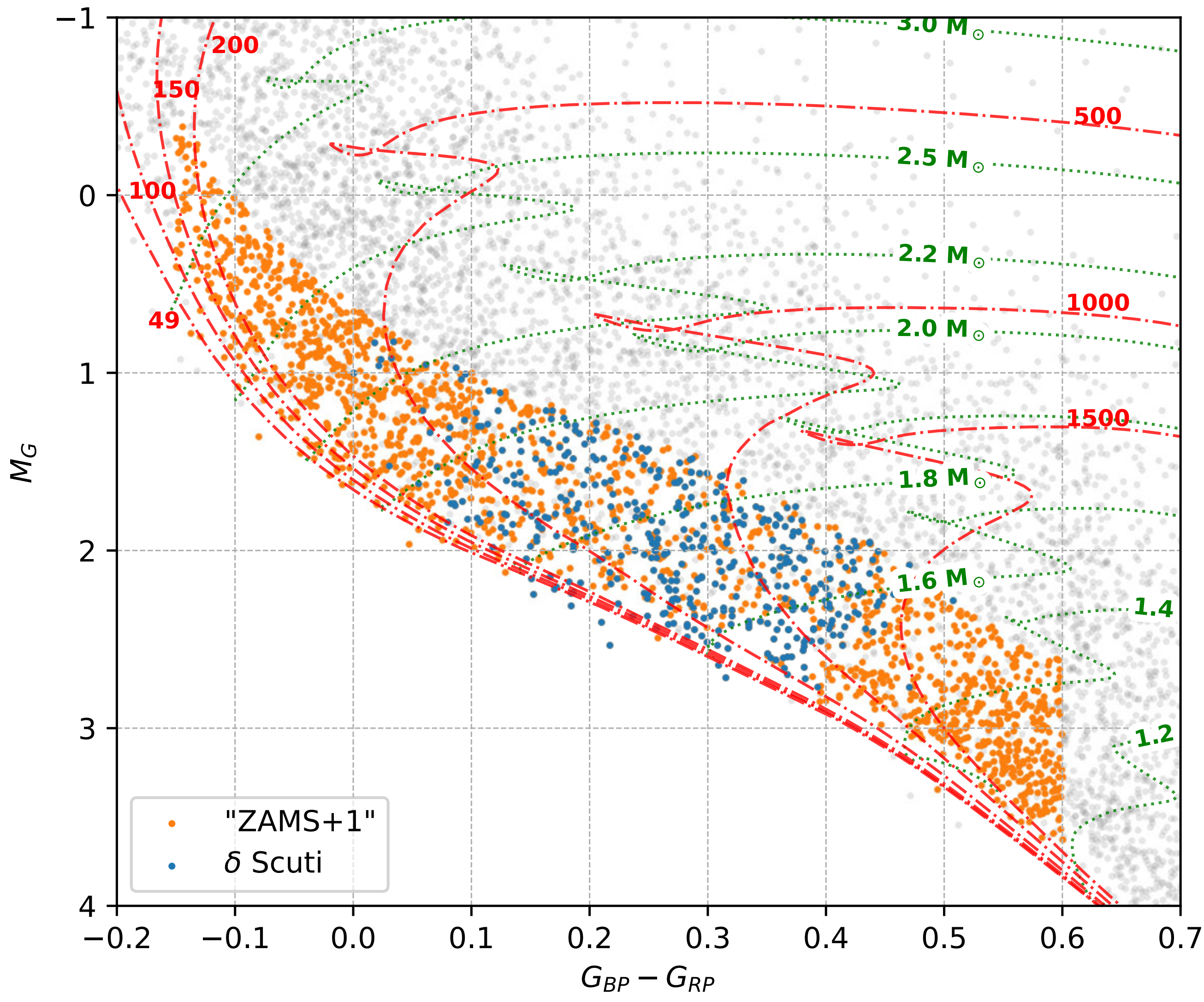

**Figure 1.** Colour–magnitude diagram for all stars with *Gaia* apparent magnitude $G < 7$. Orange and blue points constitute the 'ZAMS + 1' sample (2041 stars), where the blue points are the identified $\delta$ Scuti stars (444 stars). Isochrones (red dashed lines, in units of Myr) and evolutionary tracks (green dotted lines) are shown for solar metallicity ([Fe/H] = 0) and $v/v_{crit} = 0.4$ from MIST. This high value for rotation was chosen to better fit typically rapidly rotating $\delta$ Scuti stars (Royer, Zorec & Gómez 2007).

This paper examines a region in the colour–magnitude diagram (CMD) that is roughly orthogonal to the vertical strip studied by Read et al. (2024). We again restrict the sample to bright stars ($G < 7$) because we are particularly interested in completeness so that we can measure the true pulsator fraction. We also wished to keep the sample small enough so that each star could be inspected individually. And by choosing stars that lie within 1 mag of the zero-age main sequence (ZAMS) in the CMD, we are selecting for young stars that we can cross-match with known moving groups and associations.

In Section 2.1 we describe the selection procedure. Section 3.1 outlines our results on the fraction of stars that pulsate as $\delta$ Scuti in our sample in the instability strip. In Section 3.2, we discuss the period–luminosity relation for $\delta$ Scuti stars and explore the low frequency p-mode pulsations of some $\delta$ Scuti stars. Section 3.3 addresses the notion of completeness of our sample in terms of $\delta$ Scuti identification. Section 3.4 describes cross-matching of our sample with nearby known moving groups and associations. Appendix A highlights interesting stars and contaminants that were discarded or were difficult to label.

## 2 DATA ANALYSIS

### 2.1 Sample selection

We queried the *Gaia* archive[3] using the following filters: *Gaia* apparent magnitude $G < 7$ and $G_{BP} - G_{RP}$ in the range −0.15–0.6. The CMD is shown in Fig. 1. Absolute magnitude, $M_G$, was calculated using the relation $M_G = G + 5\log_{10}\pi - 10$, where apparent magnitude $G$ and parallax $\pi$ (in units of milliarcsec) were obtained from *Gaia* DR3 catalogue. We drew along the ZAMS a smooth curve and selected all the stars that fall within 1 mag above it, which produced a sample of 2370 stars. Owing to the high-frequency oscillations of some $\delta$ Scuti stars, we did not use *TESS* 1800-s data; we only considered 2041 stars for which *TESS* 120, 200, or 600 light curves are available, which we refer to as the 'ZAMS + 1' sample. The corresponding Nyquist frequencies are 360, 216, and 72 $\mathrm{d}^{-1}$, respectively. The 2041 stars in our sample span a distance range 12–293 pc (see Fig. 2).

[3] https://gea.esac.esa.int/archive/

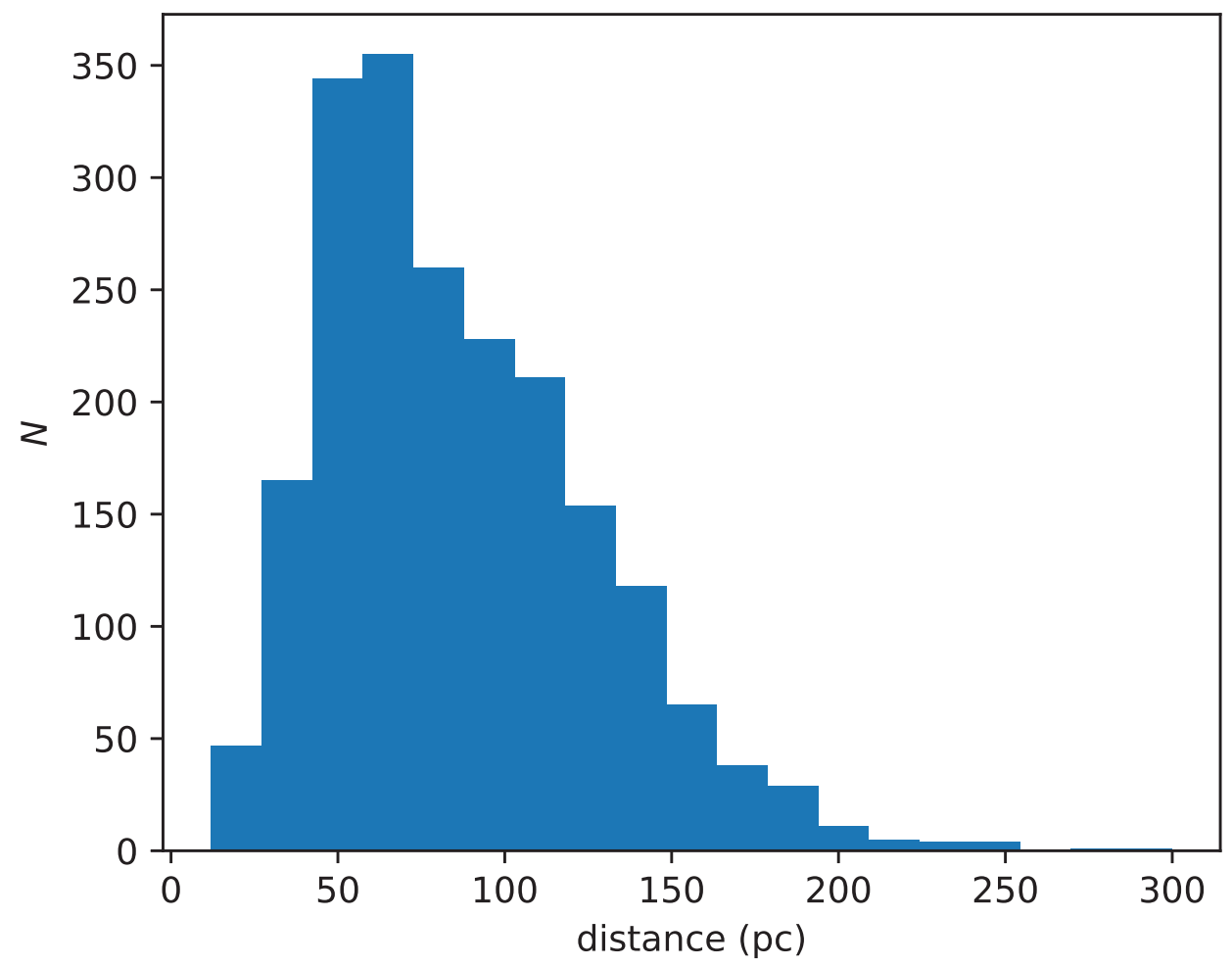

**Figure 2.** Distances of the stars in 'ZAMS + 1' sample as obtained from *Gaia* DR3 parallaxes.

Where available, we used light curves produced by the SPOC pipeline (Science Processing Operating Center; Caldwell et al. 2020). Otherwise, we used light curves from QLP (Quick Look Product; Huang et al. 2020). Both are archived at MAST[4] and we downloaded all the observations up to Sector 80. In our sample, 3.4 per cent of stars have more than 10 sectors of observation, 17.2 per cent have between 5 and 10 sectors, and the rest were observed for fewer than 5 sectors. Amplitude spectra for these light curves were computed using the Lomb–Scargle algorithm in the PYTHON LIGHTKURVE routine (lightkurve Collaboration 2018). The routine oversamples the frequency grid by a factor of 5 as default and we left this unchanged.

## 2.2 Identifying $\delta$ Scuti stars

$\delta$ Scuti stars are intermediate-mass stars with effective temperatures in the range 6000–9500 K, which broadly corresponds to $G_{BP} - G_{RP}$ in the range 0–0.6. They generally oscillate with frequencies $\geq 5\,\mathrm{d}^{-1}$, although we discuss this further in Section 3.3. We have found that the skewness of the amplitude spectra above this frequency (due to distinct, high S/N frequency peaks in broad-band noise) is useful to identify potential $\delta$ Scuti stars (Murphy et al. 2019; Hey & Aerts 2024; Read et al. 2024). However, for any given sample, the distribution of the skewness values is continuous and intermediate cases require verification (Skarka et al. 2022; Skarka & Henzl 2024). We therefore manually classified the stars in our 'ZAMS + 1' sample in the ambiguous region of skewness histogram, shown in Fig. 3. We also list several $\delta$ Scuti stars in close binary systems in Table 2 that we do not plot in our figures because the colour and brightness measured by *Gaia* for these objects cannot be confidently attributed to the $\delta$ Scuti stars in those systems. Some stars in our 'ZAMS + 1' sample have densely distributed frequency peaks in a narrow region around 10–12 $\mathrm{d}^{-1}$ (see Fig. 4), which we list in Table 1. These were described by Murphy et al. (2019, appendix B) as 'new pulsators'.

[4]https://archive.stsci.edu/missions-and-data/tess

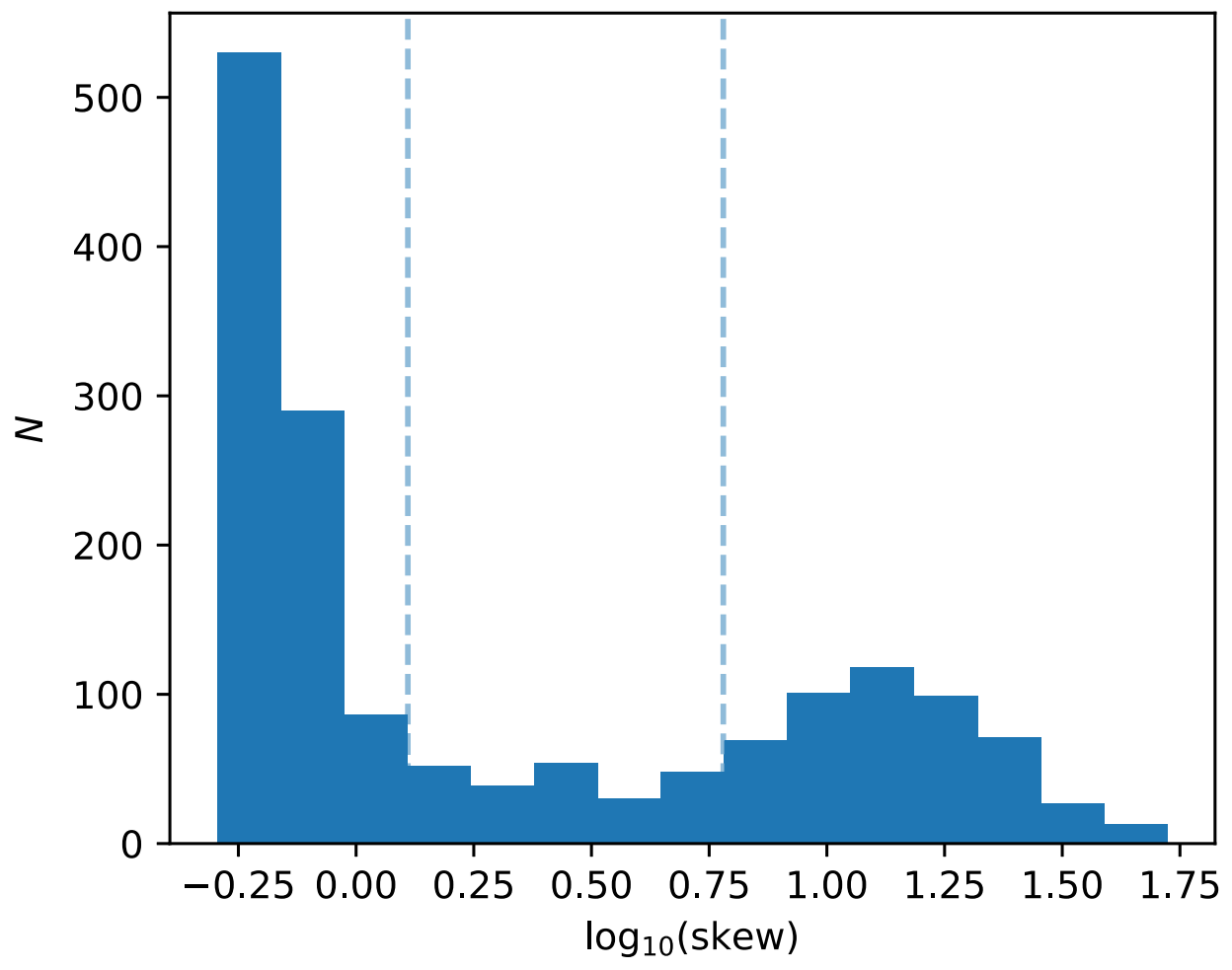

**Figure 3.** Skewness of amplitude spectra of stars in 'ZAMS + 1' sample, computed above 5 $\mathrm{d}^{-1}$. Vertical dashed lines show the range inside which we classified stars manually (see the text).

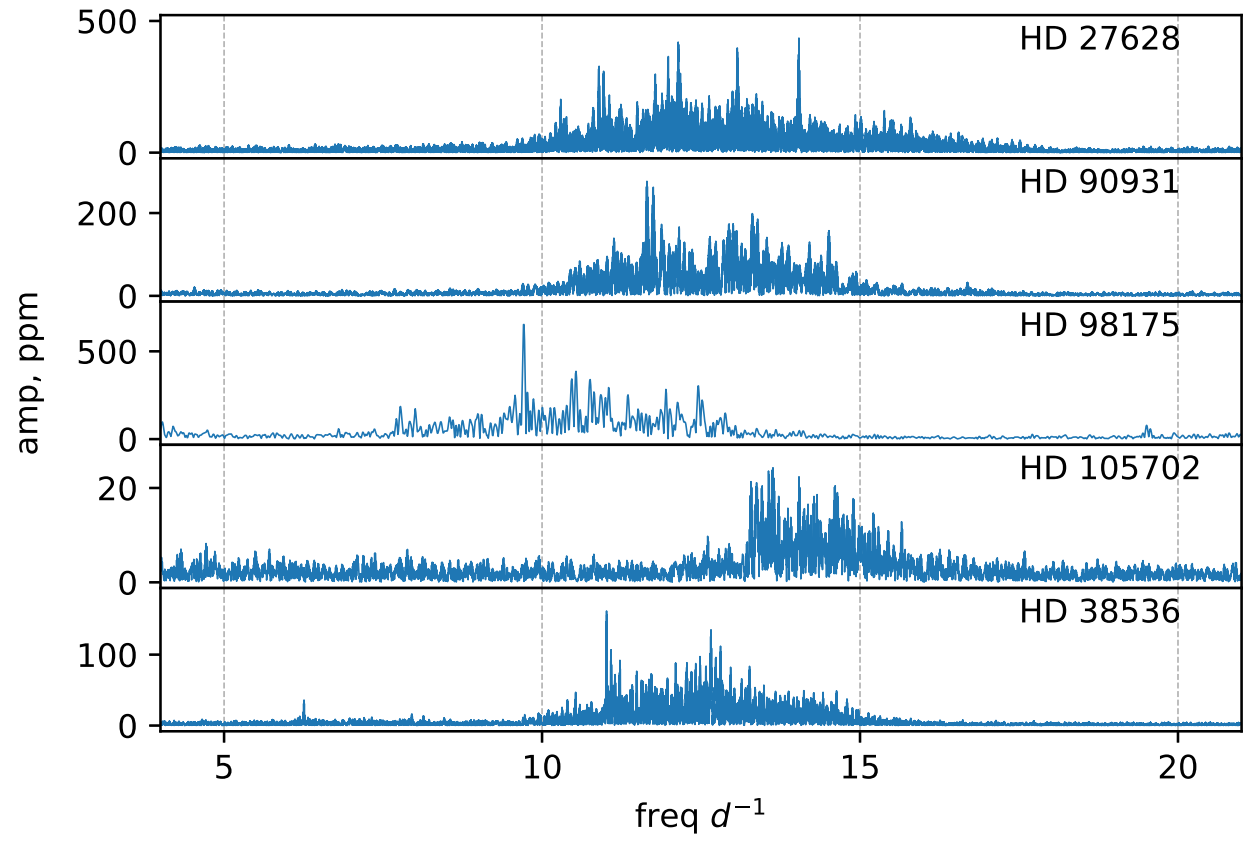

**Figure 4.** Five 'new pulsators' in 'ZAMS + 1' sample (see Section 2.2).

**Table 1.** Stars classified as 'new pulsators' (see Section 2.2).

| TIC | HD | $G$ |
|---|---|---|
| 435923755 | 27628 | 5.64 |
| 366700717 | 105702 | 5.66 |
| 100531398 | 38536 | 6.76 |
| 152056666 | 98175 | 6.78 |
| 353968091 | 90931 | 6.79 |

**Table 2.** Stars for which $\delta$ Scuti pulsations were seen in close binary systems.

| TIC | HD | $G$ |
|---|---|---|
| 160268882 | 106112 | 5.06 |
| 290409509 | 137333 | 5.54 |
| 419575080 | 110698 | 6.83 |
| 72090499 | 42116 | 6.87 |
| 95962115 | 81421 | 6.95 |

## 3 RESULTS AND DISCUSSION

### 3.1 Pulsator fraction

Within the pulsation instability strip, one might expect stars to behave similarly: oscillations, if present, would be seen in all the stars inside this region. Measurements of the fraction of star within the instability strip that show $\delta$ Scuti pulsations indicate the contrary (e.g. Guzik et al. 2014, 2015; Murphy et al. 2015, 2019; Balona & Ozuyar 2020; Bedding et al. 2023; Gootkin et al. 2024), (although Murphy et al. 2024, found close to 100 per cent pulsator fraction for the Cep-Her Complex). Read et al. (2024) speculated that the pulsation fraction might be <100 per cent possibly owing to our inability to measure oscillations with high fidelity in distant (faint) $\delta$ Scuti stars. In other words, for fainter stars photon noise might increasingly dominate signals from pulsations where they might indeed theoretically exist. The $\delta$ Scuti pulsator fraction was $\simeq$ 70 per cent for the samples in Read et al. (2024), Gootkin et al. (2024), and Bedding et al. (2023) which included stars fainter than $G > 7$, while our sample has a limit of $G < 7$. A discussion on the notion of completeness, that is 'is our sample complete enough to confidently remark upon the <100 per cent pulsator fraction?' is further outlined in Section 3.3

In Fig. 5, we see that the instability strip for our 'ZAMS + 1' sample is not 'pure'. The pulsator fraction is highest in the middle of the strip, near $G_{BP} - G_{RP} \simeq 0.3$, with $\simeq 70$ per cent and tapers off on either side. What inhibits mode excitation (at least with high enough amplitude for detection) in the remaining stars in the strip remains to be seen – at the red edge for our 'ZAMS + 1' sample, convection might dampen oscillations (Houdek 2000; Murphy et al. 2019; Gootkin et al. 2024). Our work shows no evidence of the category of stars sometimes called 'Maia variables' (Balona & Ozuyar 2020), as we do not see any $\delta$ Scuti-like pulsations bluer than $G_{BP} - G_{RP} < 0$ (see also Kahraman Aliçavuş et al. 2024).

### 3.2 Period–luminosity relation

The period–luminosity (P–L) relation for $\delta$ Scuti stars has been extensively studied (e.g. McNamara 2011; Ziaali et al. 2019; Barac et al. 2022; Martínez-Vázquez et al. 2022; Gaia Collaboration 2023; Soszyński et al. 2023; García Hernández et al. 2024; Gootkin et al. 2024; Vasigh, Ziaali & Safari 2024; Guo et al. 2025; Jia et al. 2025). In Fig. 6, we plot the absolute magnitudes of the $\delta$ Scuti stars as a function of the period of the highest amplitude mode above 5 $d^{-1}$. Note that we only show one peak per star. The P–L relation for the fundamental mode, shown as a dashed red line, was obtained by Barac et al. (2022) using a ground-based catalogue of $\delta$ Scuti stars (Rodríguez, López-González & López de Coca 2000). To better visualize this, we created a histogram in Fig. 7 of the horizontal distance of the $\delta$ Scuti stars from the Barac et al. (2022) relation. The black curve shows fixed-bandwidth kernel density estimates (KDE; Terrell & Scott 1992), which help mitigate the influence of bin size on the histogram. Many stars are seen to pulsate in the fundamental mode (excess stars near 0 distance) and around twice the period of fundamental mode (distance $\sim -0.3$). This was also seen in the samples of Ziaali et al. (2019), Jayasinghe et al. (2020), Barac et al. (2022), Soszyński et al. (2023), and Read et al. (2024). This second ridge likely corresponds to third or fourth overtone pulsators (Ziaali et al. 2019) based on arguments arising from frequency resonance with the fundamental mode. Although the aforementioned works also showed the second ridge in their P–L diagrams, we note that those samples were constructed from different stars. Most of them studied $\delta$ Scuti stars from ground-based catalogues, with preferentially higher amplitudes. These are more likely to pulsate more strongly in the fundamental mode than stars in our sample, which are preferentially younger and more likely to pulsate in higher overtones.

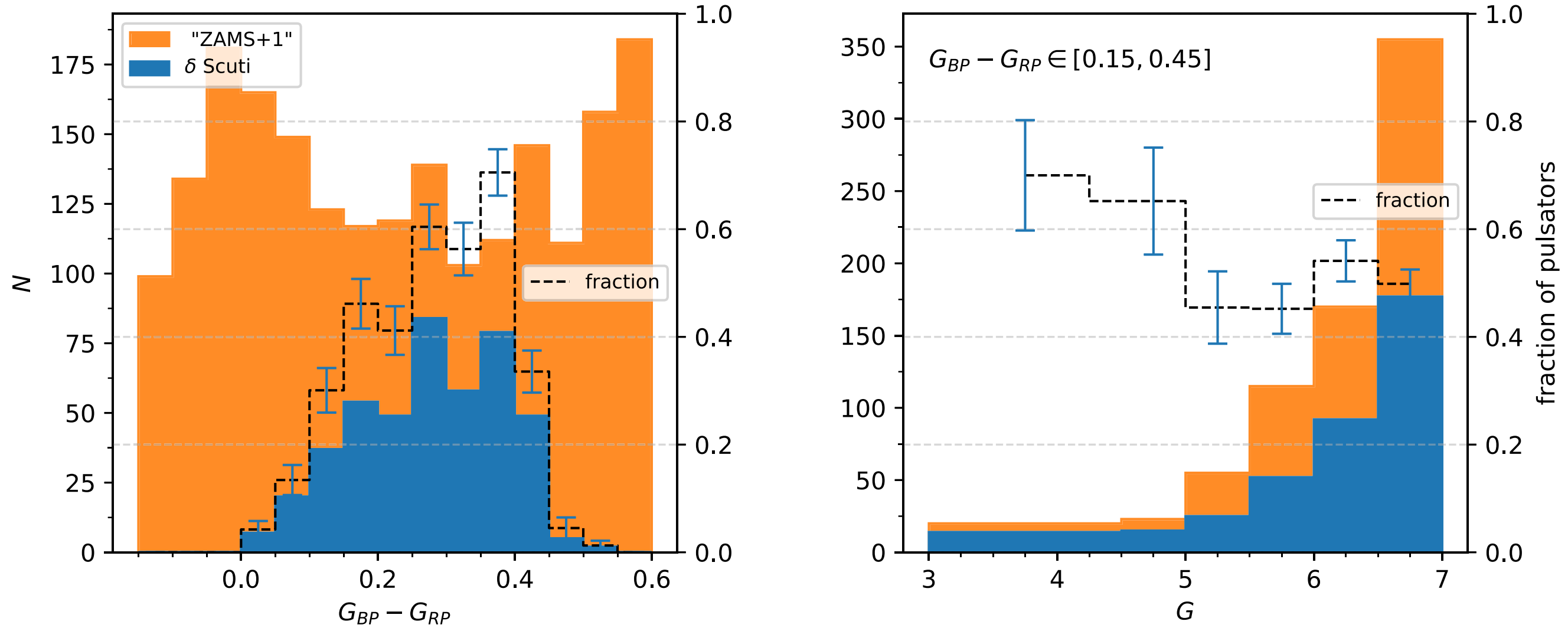

**Figure 5.** Histograms of $\delta$ Scuti stars (blue) and all stars in the 'ZAMS + 1' sample (orange), together with the pulsator fraction (dashed line). *Left*: Histograms as a function of $G_{BP} - G_{RP}$ colour. *Right*: Histograms as a function of *Gaia G* magnitude, restricted to the stars inside the instability strip ($G_{BP} - G_{RP}$ in the range 0.15–0.45).

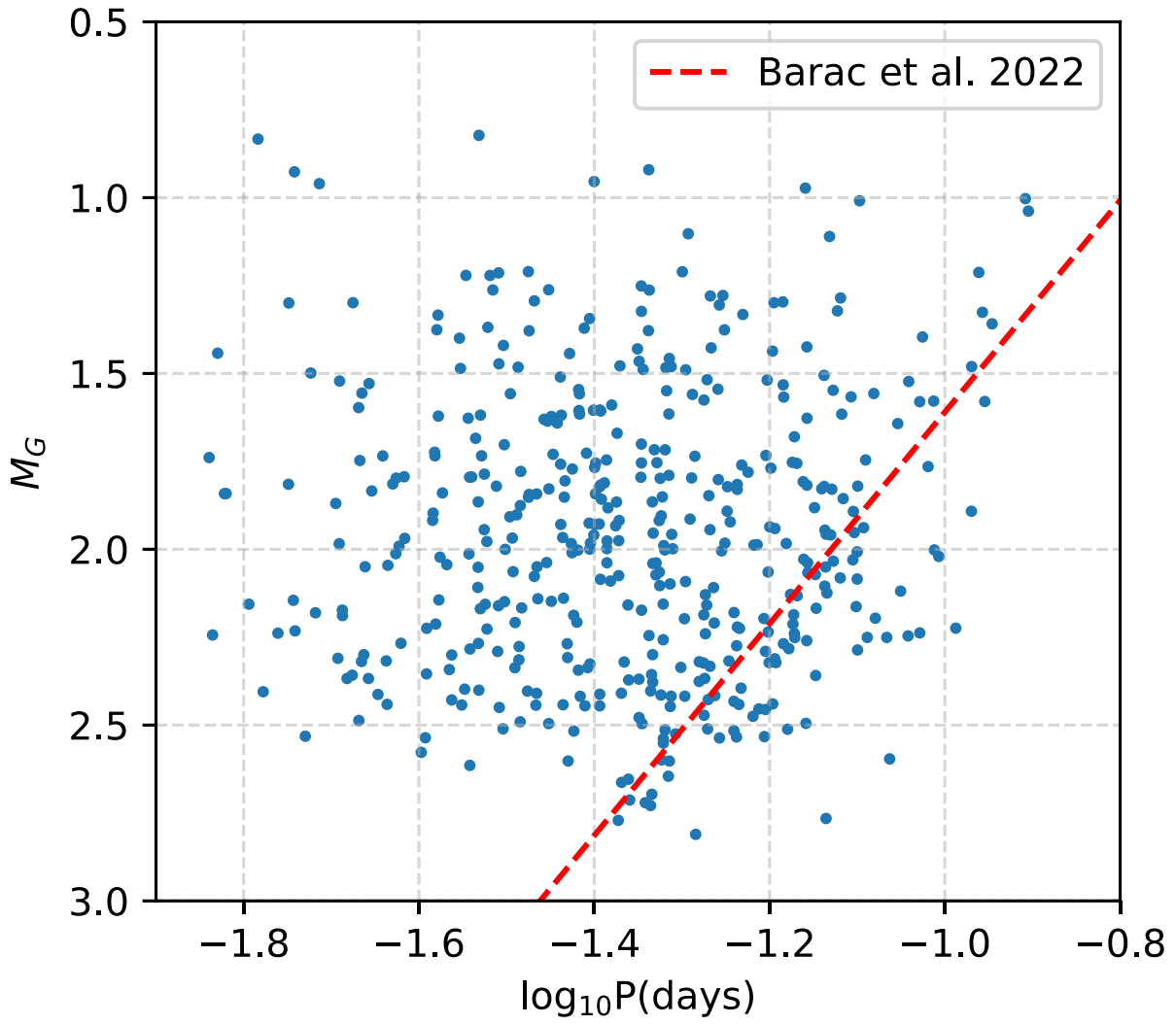

**Figure 6.** Period–luminosity plot for the $\delta$ Scuti stars, colour-coded by their $G_{BP} - G_{RP}$. Overplotted is the P–L relation obtained in Barac et al. (2022).

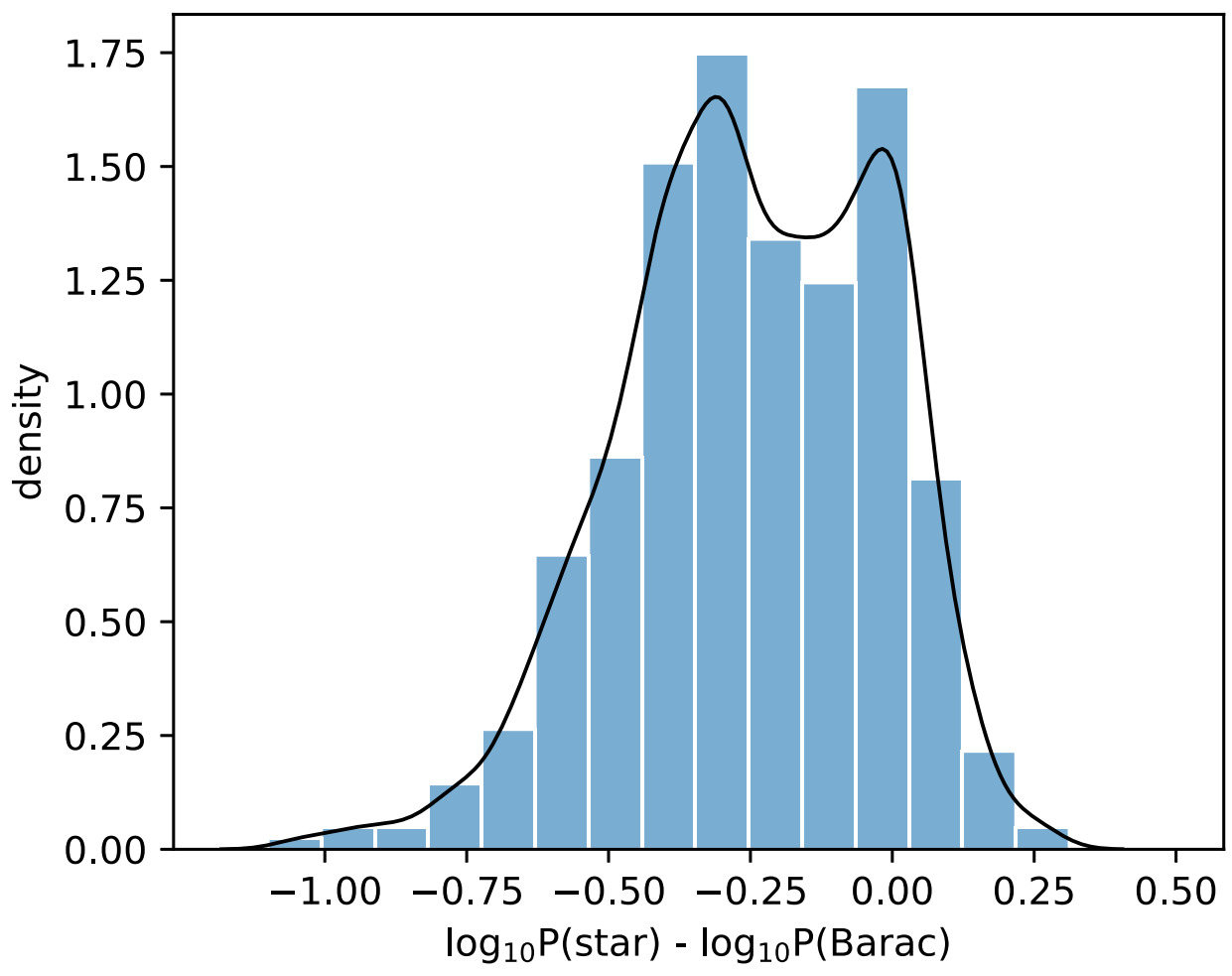

**Figure 7.** Histogram of the distance in log $P$ of the 444 $\delta$ Scuti stars from the fundamental P–L relation, calculated as the horizontal distance from the red diagonal line in Fig. 6. Solid black line is the KDE (kernel density estimate; see Section 3.2).

We note that several stars fell far to the right of the fundamental mode relation, with frequencies much lower than the fundamental p-mode frequency. These are shown in Fig. 8, with the highest peak marked by a blue circle. In most cases it is clear that this peak is not a p mode, but rather is due to low-frequency signal from rotation or g modes. For these stars, we chose to identify the highest peak about 12 $\mathrm{d}^{-1}$ (purple diamonds), and these are the peaks plotted in the P–L diagram in Fig. 6. There still remain some points to the right of the fundamental-mode P–L relation (dashed red line), and it is possible these arise from mixed modes (non-radial modes that have p- and g-mode characteristics; see Aerts et al. 2010). Although Sun-like stars only show mixed modes when they depart from the main sequence, $\delta$ Scuti stars start to show mixed modes as early as ~200 Myr (see Fig. 5; Christensen-Dalsgaard 2000).

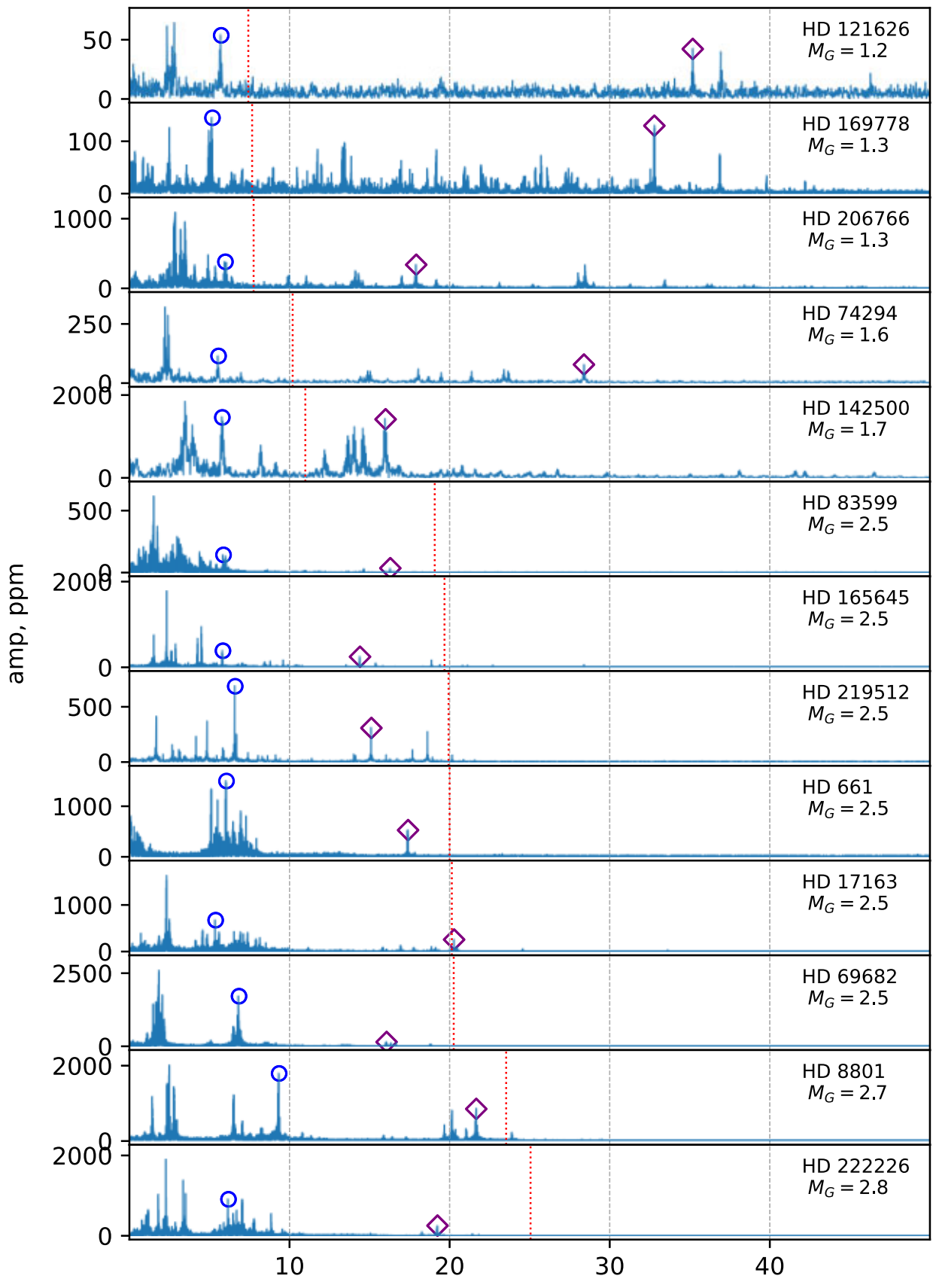

**Figure 8.** Amplitude spectra of 13 stars for which the highest-amplitude mode frequency was marked above 12 $\mathrm{d}^{-1}$ (see Section 3.2).

### 3.3 Completeness of the sample

This section addresses the question ‘How sure are we that we have identified all $\delta$ Scuti stars in our sample, in the presence of constraints such as photon noise from low-brightness stars?’ Following Read et al. (2024), in Fig. 9 we show the completeness plot. For this plot, we excluded spectra with harmonics from close binaries or rotation. The amplitudes of $\delta$ Scuti stars (blue points) typically are at least an order of magnitude above broad-band spectral noise (red points). Thus the maximum amplitude marked on the $y$-axis can hint towards stars misclassified as $\delta$ Scuti ‘not $\delta$ Scuti’. It may also help indicate contamination in measurements due to background objects (‘not $\delta$ Scuti’ stars with large amplitudes), or the presence of interesting low-amplitude $\delta$ Scuti stars, as typical amplitudes of peaks in $\delta$ Scuti spectra are on the order of a few hundred–few thousand ppm (see Bowman & Kurtz 2018).

The left panel shows the increasing trend in amplitude of ‘not $\delta$ Scuti’ stars versus $G$, which indicates an increase in white-noise level with decreasing brightness. The white noise level in the spectra also depends on the length of observation – the larger the number of sectors, the lesser the white noise, since more observation frequency bins are available over which to distribute broad-band noise content. We often find that on the cooler side of the instability strip, leaked spectral power from lower frequency g modes (or rotational peaks, Rossby modes, etc.) are misidentified as noise peaks. To this end, we performed three adjustments to the points belonging to the 'not $\delta$ Scuti' stars in the right panel of Fig. 9:

(i) We considered peaks above 12 $d^{-1}$ (rather than 5 $d^{-1}$), minimizing the chances of misidentifying leaked spectral power from lower frequency variability as noise peaks.

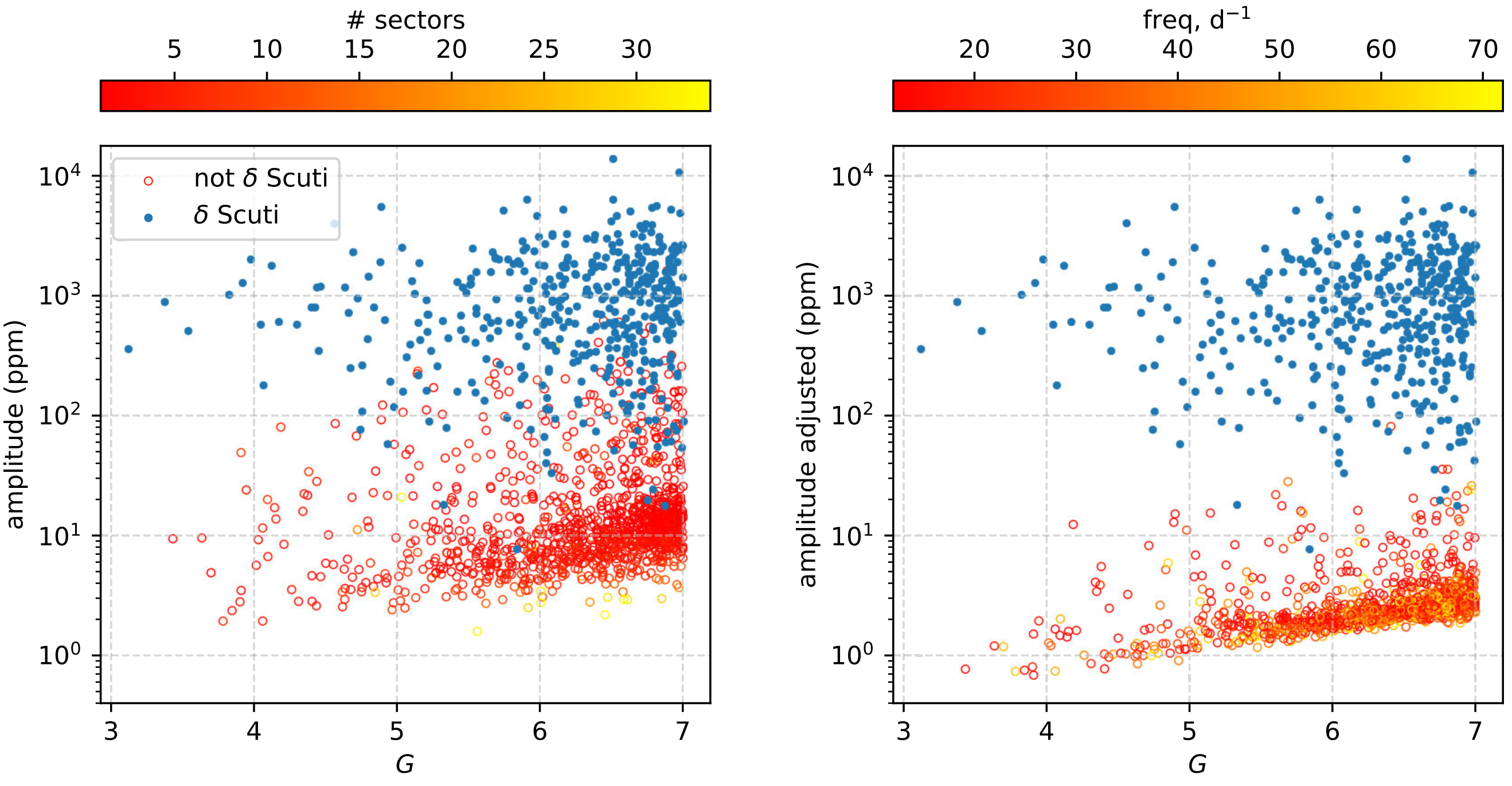

**Figure 9.** *Left*: Highest-amplitude peak above 5 $d^{-1}$ versus $G$ for $\delta$ Scuti stars and 'not $\delta$ Scuti' in 'ZAMS +1' sample. Spectra containing harmonics (binaries or rotation) are not shown. *Right*: Same as left, except the amplitudes for 'not $\delta$ Scuti' (1) are adjusted for length of observation, and (2) average of the entire spectra above 12 $d^{-1}$ is calculated instead of single highest peak above 5 $d^{-1}$ (see Section 3.3).

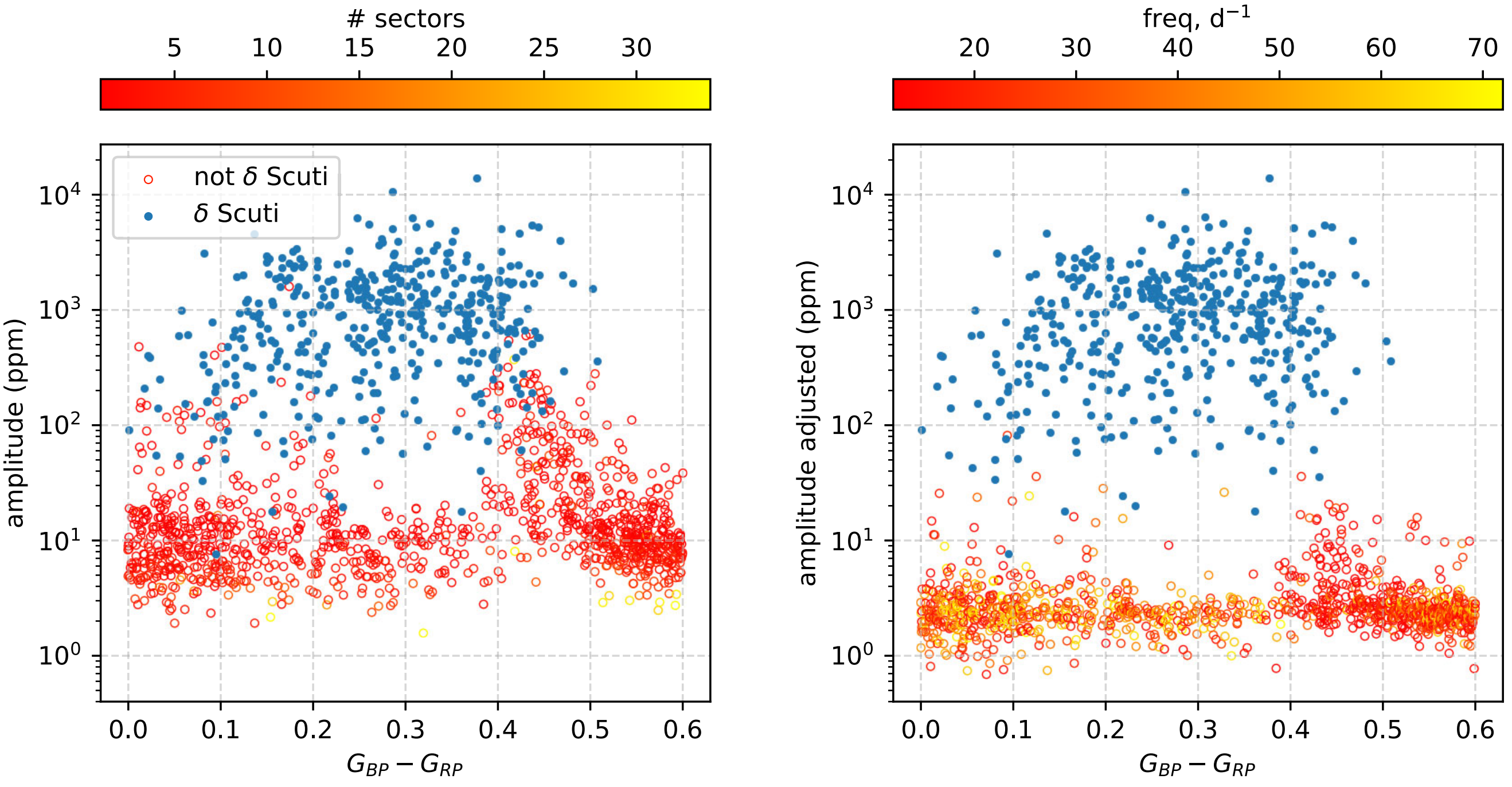

**Figure 10.** *Left*: Highest-amplitude peak above 5 $d^{-1}$ versus $G_{BP} - G_{RP}$ for $\delta$ Scuti stars and 'not $\delta$ Scuti' in 'ZAMS + 1' sample. Spectra containing harmonics (binaries or rotation) are not shown. *Right*: Same as left, except the amplitudes for 'not $\delta$ Scuti' (1) are adjusted for length of observation, and (2) average of the entire spectra above 12 $d^{-1}$ is calculated instead of single highest peak above 5 $d^{-1}$ (see Section 3.3).

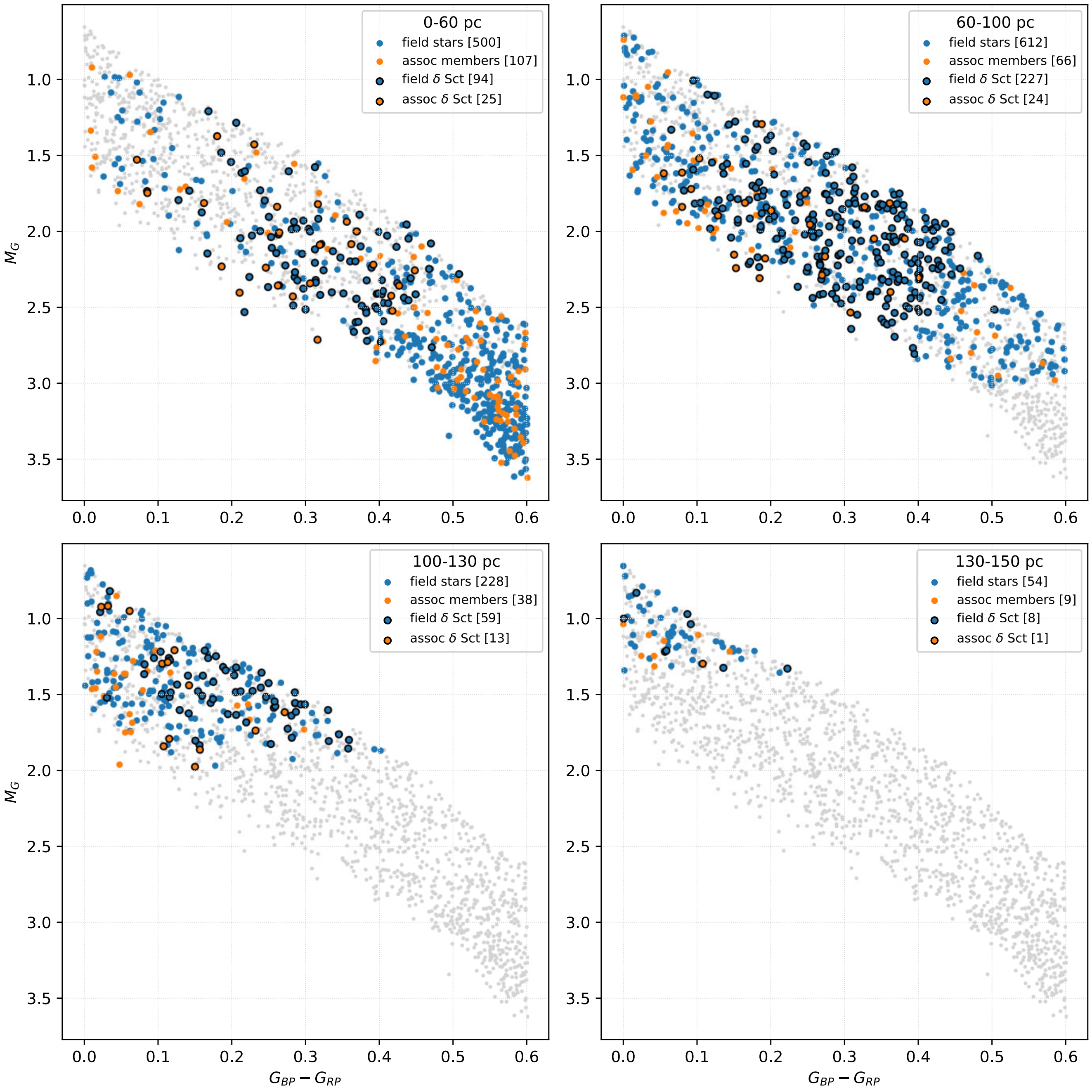

**Figure 11.** Colour–magnitude diagrams for the examined sub-sample described in Section 3.4. The stars are colour-coded according to their membership. In each panel, we plot stars selected by distance, on top of the whole sub-sample, which we marked with grey dots.

(ii) Instead of considering only the highest peak, we considered the average of spectra, as the average behaviour is more meaningful than the highest amplitude peak in a white noise region.

(iii) To adjust for the effect of different observation durations for different stars, we multiplied the highest amplitude by $\sqrt{N}/\mathrm{Med}(N_{\mathrm{noise}})$ where $N$ is the number of sectors for a particular 'not $\delta$ Scuti' and $\mathrm{Med}(N_{\mathrm{noise}})$ is the median of all the 'not $\delta$ Scuti' sectors.

As brightness decreases ($G$ increases), amplitudes of $\delta$ Scuti stars and 'not $\delta$ Scuti' stars become similar, and we may miss the identification of $\delta$ Scuti stars simply because they are faint. This can be arguably seen as $G$ approaches 5 and fainter.

We show a similar plot in Fig. 10 but with $G_{BP} - G_{RP}$ on the $x$-axis. There is a relative underdensity of points within the instability strip ($G_{BP} - G_{RP} \simeq 0.3$) which appears to be a 'forbidden zone', suggesting that a star can possibly be a $\delta$ Scuti with relatively high amplitude or a low amplitude 'not $\delta$ Scuti'.

**Table 3.** Table values are obtained by running the BANYAN Σ code (Gagné et al. 2018) on a subsample of the initial one. Distances are average values from MOCA data base.

| Association Full name | Association Acronym | 'ZAMS + 1' [a] stars | 'ZAMS + 1' [a] $\delta$ Scuti | Fraction of $\delta$ Scuti stars | Distance (pc) | Age (Myr) |
|---|---|---|---|---|---|---|
| 118 Tau | 118TAU | 1 | 0 | 0 | 103 | $\sim$10 [(1)] |
| AB Doradus | ABDMG | 20 | 6 | 0.30 | 42 | $133^{+15}_{-20}$ [(1)] |
| Argus | ARG | 51 | 13 | 0.25 | 56 | 45 – 50 [(2)] |
| $\beta$ Pictoris | BPMG | 16 | 5 | 0.31 | 46 | $26 \pm 3$ [(3)] |
| Carina | CAR | 3 | 1 | 0.33 | 76 | 45 [(4)] |
| Carina-Near | CARN | 9 | 1 | 0.11 | 33 | 200 [(5)] |
| Coma Benerices | CBER | 8 | 4 | 0.50 | 86 | 562 [(6)] |
| Columba | COL | 7 | 3 | 0.43 | $\sim$69 | 42 [(4)] |
| Corona Australis | CRA | 0 | 0 | 0 | 148 | 4 – 5 [(7)] |
| $\epsilon$ Chamaleontis | EPSC | 2 | 2 | 1 | 96 | 3.7 [(8)] |
| $\eta$ Chamaleontis | ETAC | 0 | 0 | 0 | 99 | 6.5 [(8)] |
| Hyades cluster | HYA | 39 | 9 | 0.23 | 47 | $695^{+85}_{-67}$ [(9)] |
|  | IC2391 | 0 | 0 | 0 | 151 | $57.7^{+0.5}_{-1.0}$ [(9)] |
|  | IC2602 | 0 | 0 | 0 | 150 | $52.5^{+2.2}_{-3.7}$ [(9)] |
| Lower Centaurus Crux | LCC | 12 | 7 | 0.58 | 110 | 15 [(10)] |
| $\mu$ Tau | MUTAU | 0 | 0 | 0 | 148 | 61 [(11)] |
| Octans | OCT | 0 | 0 | 0 | 129 | 30 – 40 [(12)] |
| Pisces-Eridanus | PERI | 9 | 3 | 0.33 | 115 | $\sim$120 [(13)] |
| Platais 8 | PL8 | 0 | 0 | 0 | 133 | $\sim$60 [(14)] |
| Pleiades cluster | PLE | 6 | 0 | 0 | 128 | $127.4^{+6.3}_{-10}$ [(9)] |
| Praesepe cluster | PRAE | 0 | 0 | 0 | 185 | 617 [(15)] |
| $\rho$ Ophiuchi | ROPH | 0 | 0 | 0 | 123 | $< 2$ [(16)] |
| Taurus-Auriga | TAU | 3 | 0 | 0 | 125 | 1 – 2 [(17)] |
| Tucana-Horologium | THA | 5 | 1 | 0.20 | 50 | $51.0^{+0.5}_{-0.2}$ [(9)] |
| 32 Orionis | THOR | 2 | 0 | 0 | 91 | $25.0^{+0.7}_{-0.8}$ [(9)] |
| TW Hydrae | TWA | 2 | 1 | 0.50 | 59 | 10 [(4)] |
| Upper Centaurus Lupus | UCL | 17 | 4 | 0.23 | 131 | 16 [(10)] |
| Upper CrA | UCRA | 0 | 0 | 0 | 149 | $\sim$10 [(1)] |
| Ursa Major cluster | UMA | 3 | 1 | 0.33 | 24 | 414 [(18)] |
| Upper Scorpius | USCO | 0 | 0 | 0 | 114 | 10 [(10)] |
| Volans-Carina | VCA | 2 | 0 | 0 | 87 | 89 [(1)] |
| $\chi$ Fornacis | XFOR | 3 | 2 | 0.67 | 103 | $\sim$40 [(19)] |

*Notes.* [a] We performed the analysis selecting a subsample of 1614 stars out of the original 2041. See Section 3.4 for more details.

*References.* (1) Gagné et al. (2018), (2) Zuckerman (2019), (3) Malo et al. (2014), (4) Bell, Mamajek & Naylor (2015), (5) Zuckerman et al. (2006), (6) Silaj & Landstreet (2014), (7) Gennaro, Prada Moroni & Tognelli (2012), (8) Murphy, Lawson & Bessell (2013), (9) Galindo-Guil et al. (2022), (10) Pecaut & Mamajek (2016), (11) Gagné et al. (2020), (12) Murphy & Lawson (2015), (13) Curtis et al. (2019), (14) Platais, Kozhurina-Platais & van Leeuwen (1998), (15) Gossage et al. (2018), (16) Wilking, Gagné & Allen (2008), (17) Kenyon & Hartmann (1995), (18) Jones et al. (2015), and (19) Zuckerman, Klein & Kastner (2019).

### 3.4 Cross-matching with nearby young associations

Stars belonging to the A spectral class tend to have a short lifespan. They are often found within young stellar populations, such as stellar associations of coeval stars, where they formed from the same dense molecular cloud, and are mostly younger than a few hundred million years. We expect some of the young $\delta$ Sct pulsators, which are A-type stars, to be found within young stellar associations like OB associations and open clusters.

Bedding et al. (2023) examined a narrow range of A- and F-type stars within the Pleiades open cluster, searching for pulsators. Moreover, a search for pulsators with regular pulsation patterns performed by Bedding et al. (2020) indicated that some were found within associations. Addressing these stars does not only give an insightful characterization of pulsators within young nearby associations, but it is also applicable to more distant structures, as Murphy et al. (2024) did for the Cep-Her complex ($\sim$320 pc distant). Connecting the pulsator fraction to the associations' properties has potential interest within the field of asteroseismology.

We cross-matched the sample with the list of known young nearby associations, using BANYAN Σ (Bayesian Analysis for Nearby Young AssociatioNs) algorithm (Gagné et al. 2018). This tool is based on Bayesian inference and it has been implemented to calculate, given a sample of stars, their membership probabilities to bona fide members of clusters and associations as far as 150 pc from the Sun. Currently, BANYAN Σ includes kinematic models for 32 associations, namely the 27 nearby associations described in Gagné et al. (2018), together with the well-studied Praesepe open cluster, and the recently discovered Volans–Carina, Argus, $\mu$ Tau, Pisces–Eridanus moving groups. Each stellar association is modelled with multivariate Gaussians in a six-dimensional XYZUVW space, working with Galactic position (XYZ) and space velocity (UVW). The classification algorithm requires direct kinematic observables for the examined stars, and then it determines the probability that they belong to either the Galactic field or one young nearby association. Sky coordinates and proper motion of the stars are the only required input parameters, however, to increase accuracy and reliability of the output membership probabilities, radial velocity and distance can be included. For our analysis, we used sky position, proper motion and radial velocities as inputs, and after running the code, we selected only the stars exhibiting membership probabilities above 90 per cent. These values were then cross-matched with the $\delta$ Scuti stars presented in the previous sections. Table 3 reports how many stars we can confidently say are members of young nearby associations, how many of them are $\delta$ Scuti pulsators, and we report the corresponding association ages and distances from the Montreal Open Clusters and Associations (MOCA) data base.

The analysis is performed on a sub-sample of the original one. We address stars up to the distance 150 pc, and within the colour range $BP–RP = 0–0.6$, for a total of 1614 stars out of the initial 2041. The four CMDs in Fig. 11 provide a useful overview of the sub-sample distribution. Having determined the membership probability for the stars with BANYAN $\Sigma$, we sorted the stars by distance and selected four distance bins. Then, we distinguished between association members and field stars. BANYAN $\Sigma$ is currently limited to the maximum distance of 150 pc, therefore, we are able to estimate how many stars are members of these listed associations only. Next, we are interested in the fraction of $\delta$ Scuti stars among them. One key finding is that ∼13.6 per cent of the stars (220 out of 1614) within 150 pc are probable members of one of the young nearby associations listed. In addition, most of the identified $\delta$ Scuti stars lie between 60 and 100 pc. Overall, $\delta$ Scuti stars from our sample account for ∼29 per cent of stars (63 out of 220) that are members of nearby associations. This implies that $\delta$ Scuti stars in these associations comprise ∼3 per cent of our total examined sample.

## 4 SUMMARY

With the help of light curves from *TESS* and *Gaia* DR3 measurements for stars lying near the ZAMS, our goal was to characterize bright, young $\delta$ Scuti stars.

We explored various pulsation characteristics of $\delta$ Scuti stars such as fraction of pulsators, the overtones of pulsation, and outlined the observed the lower limits in frequency and amplitude of $\delta$ Scuti oscillations in our 'ZAMS + 1' sample. We found that in the middle of the instability strip, around 70 per cent of stars pulsate in p modes which we associate with $\delta$ Scuti, similar to Murphy et al. (2019), Bedding et al. (2023), and Read et al. (2024). We find $\delta$ Scuti pulsation amplitudes as low as $\simeq 8$ ppm (see Fig. 9); although *Kepler* predominantly observed distant, relatively fainter $\delta$ Scuti stars, its longer observation duration facilitated in measuring pulsation amplitude with high fidelity by helping to reduce white noise levels. Murphy et al. (2019) used *Kepler* long-cadence data and thus found $\delta$ Scuti pulsation amplitudes as low as 10 ppm. Whether $\delta$ Scuti stars can pulsate with even lower amplitude remains an interesting open question.

We found some $\delta$ Scuti stars that displayed unusually low frequency peaks. The fundamental p-mode frequency, which is the lowest oscillation acoustic frequency for a $\delta$ Scuti star, should place it close to the Barac et al. (2022) red-dashed line in Fig. 6. Some stars fell to the bottom right of this P–L relation, raising questions about the lowest frequency a pressure mode can oscillate in given a star's size. It is possible that brightness measurements may be inaccurate for some of such stars, leading to them being placed incorrectly in the P–L diagram. The presence of mixed modes may also explain why oscillations in $\delta$ Scuti stars are observed below the theoretical fundamental p-mode frequencies. Rapid rotators seen edge-on can also place stars to the right of the fundamental mode relation, as surface rotation affects measured luminosities.

Finally, we cross-matched our sample to understand what fraction of them belonged to nearby young associations. We found that 63 $\delta$ Scuti stars from our sample were members of young nearby association. The BANYAN $\Sigma$ code is designed to find association members upto a distance of 150 pc; for our sample we do not suspect cross-matching results to change drastically when including distant associations. This is because the sample we analysed, based on isochrones (MESA Isochrones & Stellar Tracks (MIST); see Fig. 1), contains stars ($\delta$ Scuti) older than 1 Gyr – associations become gravitationally unbound in a few hundred Myr.

This work looked at a sample that was approximately orthogonal in the CMD to the Read et al. (2024) sample; future work should consider a whole sky survey of $\delta$ Scuti stars with *TESS*. Based on our discussion of the completeness plot (Fig. 9), we also wish to note that a thorough investigation of the amplitude lower limits of $\delta$ Scuti pulsations should reveal interesting properties of mode driving mechanisms, which future models must incorporate.

## ACKNOWLEDGEMENTS

We gratefully acknowledge support from the Australian Research Council through Future Fellowship FT210100485, and Laureate Fellowship FL220100117. This work has made use of data from the European Space Agency (ESA) mission *Gaia*, (https://www.cosmos.esa.int/gaia), processed by the *Gaia* Data Processing and Analysis Consortium (DPAC, https://www.cosmos.esa.int/web/gaia/dpac/⟨?Undefined?⟩⟨?PMU ?⟩⟨?Undefined?⟩consortium). This research made use of the Montreal Open Clusters and Associations (MOCA) data base (https://mocadb.ca), operated at the Montréal Planétarium (Gagné et al., in preparation). Funding for the DPAC has been provided by national institutions, in particular the institutions participating in the *Gaia* Multilateral Agreement. We are grateful to the entire *Gaia* and *TESS* teams for providing the data used in this paper. This work made use of several publicly available PYTHON packages: ASTROPY (Astropy Collaboration 2013, 2018), LIGHTKURVE (Lightkurve Collaboration 2018), MATPLOTLIB (Hunter 2007), NUMPY (Harris et al. 2020), and SCIPY (Virtanen et al. 2020).

## DATA AVAILABILITY

The *TESS* data underlying this article are available at the MAST Portal (Barbara A. Mikulski Archive for Space Telescopes), at https://mast.stsci.edu/portal/Mashup/Clients/Mast/Portal.html.

## REFERENCES

Aerts C., 2015, Astron. Nachr., 336, 477
Aerts C., 2021, Rev. Mod. Phys., 93, 015001
Aerts C., Christensen-Dalsgaard J., Kurtz D. W., 2010, Asteroseismology. Springer, Berlin
Antoci V. et al., 2014, ApJ, 796, 118
Antoci V. et al., 2019, MNRAS, 490, 4040
Astropy Collaboration, 2013, A&A, 558, A33
Astropy Collaboration, 2018, AJ, 156, 123
Balona L. A., Ozuyar D., 2020, MNRAS, 493, 5871
Balona L. A., Daszyńska-Daszkiewicz J., Pamyatnykh A. A., 2015, MNRAS, 452, 3073
Barac N., Bedding T. R., Murphy S. J., Hey D. R., 2022, MNRAS, 516, 2080
Barceló Forteza S., Moya A., Barrado D., Solano E., Martín-Ruiz S., Suárez J. C., García Hernández A., 2020, A&A, 638, A59
Bedding T. R. et al., 2020, Nature, 581, 147
Bedding T. R. et al., 2023, ApJ, 946, L10

Bell C. P. M., Mamajek E. E., Naylor T., 2015, MNRAS, 454, 593
Bowman D. M., Kurtz D. W., 2018, MNRAS, 476, 3169
Bowman D. M., Kurtz D. W., Breger M., Murphy S. J., Holdsworth D. L., 2016, MNRAS, 460, 1970
Caldwell D. A. et al., 2020, Res. Notes Am. Astron. Soc., 4, 201
Christensen-Dalsgaard J., 2000, in Breger M., Montgomery M., eds, ASP Conf. Ser. Vol. 210, Delta Scuti and Related Stars. Astron. Soc. Pac., San Francisco, p. 187
Cox J. P., 1980, Theory of Stellar Pulsation, Vol. 2. Princeton Univ. Press, Princeton
Curtis J. L., Agüeros M. A., Mamajek E. E., Wright J. T., Cummings J. D., 2019, ApJ, 158, 77
Dholakia S., Murphy S. J., Huang C. X., Venner A., Wright D., 2025, MNRAS, 536, 2313
Fritzewski D. J., Van Reeth T., Aerts C., Van Beeck J., Gossage S., Li G., 2024, A&A, 681, A13
Gagné J. et al., 2018, ApJ, 856, 35
Gagné J., David T. J., Mamajek E. E., Mann A. W., Faherty J. K., Bédard A., 2020, ApJ, 903, 30
Gaia Collaboration et al., 2023, A&A, 674, A36
Galindo-Guil F. J. et al., 2022, A&A, 664, A70
García R. A., Ballot J., 2019, Living Rev. Sol. Phys., 16, 4
García Hernández A., Pascual-Granado J., Lares-Martiz M., Mirouh G. M., Suárez J. C., Barceló Forteza S., Moya A., 2024, in de Grijs R., Whitelock P. A., Catelan M., eds, Proc. IAU Symp. 376, At the Crossroads of Astrophysics and Cosmology: Period-luminosity Relations in the 2020s. Cambridge Univ. Press, Cambridge, p. 239
Gennaro M., Prada Moroni P. G., Tognelli E., 2012, MNRAS, 420, 986
Gootkin K., Hon M., Huber D., Hey D. R., Bedding T. R., Murphy S. J., 2024, ApJ, 972, 137
Gossage S., Conroy C., Dotter A., Choi J., Rosenfield P., Cargile P., Dolphin A., 2018, ApJ, 863, 18
Guo Fangzhou, Bloom Joshua S., Wang Xiaofeng, Chen Liyang, Lin Jie, Chen Xiaodian, Mo Jun, Zhang Jicheng, Yan Shengyu, Liu Qichun, Peng Haowei, Jiang Xiaojun, Ma Xiaoran, Xiang Danfeng, Wenxiong Li, , 2025,AAP, 699, A115
Guzik J. A., 2021, Front. Astron. Space Sci., 8, 55
Guzik J. A., Bradley P. A., Jackiewiecz J., Uytterhoeven K., Kinemuchi K., 2014, Astron. Rev., 9, 41
Guzik J. A., Bradley P. A., Jackiewicz J., Molenda-Zakowicz J., Uytterhoeven K., Kinemuchi K., 2015, Astron. Rev., 11, 1
Guzik J. A., Fontes C. J., Fryer C., 2018, Atoms, 6, 31
Harris C. R. et al., 2020, Nature, 585, 357
Hey D., Aerts C., 2024, A&A, 688, A93
Hey D. R., Montet B. T., Pope B. J. S., Murphy S. J., Bedding T. R., 2021, AJ, 162, 204
Houdek G., 2000, in Breger M., Montgomery M., eds, ASP Conf. Ser. Vol. 210, Delta Scuti and Related Stars. Astron. Soc. Pac., San Francisco, p. 454
Huang C. X. et al., 2020, Res. Notes Am. Astron. Soc., 4, 204
Huber D. et al., 2012, ApJ, 760, 32
Hunter J. D., 2007, Comput. Sci. Eng., 9, 90
Jayasinghe T. et al., 2020, MNRAS, 493, 4186
Jia Q., Chen X., Wang S., Deng L., Zhang J., Jiang Q., 2025, ApJ, 984, 89
Jones J. et al., 2015, Am. Astron. Soc, 225, 112
Kahraman Aliçavuş F., Handler G., Chowdhury S., Niemczura E., Jayaraman R., De Cat P., Ozuyar D., Aliçavuş F., 2024, PASA, 41, e082
Kenyon S. J., Hartmann L., 1995, ApJS, 101, 117
Kurtz D., 2022, ARA&A, 60, 31
Lebreton Y., Goupil M. J., 2014, A&A, 569, A21
Lightkurve Collaboration, 2018, Astrophysics Source Code Library, record ascl:1812.013
Malo L., Doyon R., Feiden G. A., Albert L., Lafreniére D., Artigau E., Gagné J., Riedel A., 2014, ApJ, 792, 37
Martínez-Vázquez C. E., Salinas R., Vivas A. K., Catelan M., 2022, ApJ, 940, L25
McNamara D. H., 2011, AJ, 142, 110
Moya A., Suárez J. C., García Hernández A., Mendoza M. A., 2017, MNRAS, 471, 2491
Murphy S., Lawson W. A., 2015, MNRAS, 447, 1267
Murphy S., Lawson W. A., Bessell M. S., 2013, MNRAS, 435, 1325
Murphy S. J., Bedding T. R., Niemczura E., Kurtz D. W., Smalley B., 2015, MNRAS, 447, 3948
Murphy S. J., Hey D., Van Reeth T., Bedding T. R., 2019, MNRAS, 485, 2380
Murphy S. J., Paunzen E., Bedding T. R., Walczak P., Huber D., 2020a, MNRAS, 495, 1888
Murphy S. J., Saio H., Takada-Hidai M., Kurtz D. W., Shibahashi H., Takata M., Hey D. R., 2020b, MNRAS, 498, 4272
Murphy S. J., Joyce M., Bedding T. R., White T. R., Kama M., 2021, MNRAS, 502, 1633
Murphy S. J., Bedding T. R., White T. R., Li Y., Hey D., Reese D., Joyce M., 2022, MNRAS, 511, 5718
Murphy S. J., Bedding T. R., Gautam A., Kerr R. P., Mani P., 2024, MNRAS, 534, 3022
Olmschenk G. et al., 2024, AJ, 168, 83
Pamos Ortega D., García Hernández A., Suárez J. C., Pascual Granado J., Barceló Forteza S., Rodón J. R., 2022, MNRAS, 513, 374
Pamos Ortega D., Mirouh G. M., García Hernández A., Suárez Yanes J. C., Barceló Forteza S., 2023, A&A, 675, A167
Pecaut M. J., Mamajek E. E., 2016, MNRAS, 461, 794
Platais I., Kozhurina-Platais V., van Leeuwen F., 1998, ApJ, 116, 2423
Read A. K. et al., 2024, MNRAS, 528, 2464
Ricker G. R. et al., 2015, J. Astron. Telesc. Instrum. Syst., 1, 014003
Riello M. et al., 2021, A&A, 649, A3
Rodríguez-Martín J. E., García Hernández A., Suárez J. C., Rodón J. R., 2020, MNRAS, 498, 1700
Rodríguez E., López-González M. J., López de Coca P., 2000, A&AS, 144, 469
Royer F., Zorec J., Gómez A. E., 2007, A&A, 463, 671
Sánchez Arias J. P., Córsico A. H., Althaus L. G., 2017, A&A, 597, A29
Silaj J., Landstreet J. D., 2014, A&A, 566, 18
Skarka M., Henzl Z., 2024, A&A, 688, A25
Skarka M. et al., 2022, A&A, 666, A142
Soszyński I. et al., 2023, AcA, 73, 105
Steindl T., Zwintz K., Bowman D. M., 2021, A&A, 645, A119
Stellingwerf R. F., 1979, ApJ, 227, 935
Streamer M., Ireland M. J., Murphy S. J., Bento J., 2018, MNRAS, 480, 1372
Terrell G. R., Scott D. W., 1992, Ann. Stat., 20, 1236
Vasigh F., Ziaali E., Safari H., 2024, ApJ, 969, 19
Virtanen P. et al., 2020, Nat. Methods, 17, 261
Wilking B. A., Gagné M., Allen L. E., 2008, in Bo Reipurth, ed, Handbook of Star Forming Regions, Volume II: The Southern Sky ASP Monograph Publications. Vol. 5, Astron. Soc. Pac., p. 351
Ziaali E., Bedding T. R., Murphy S. J., Van Reeth T., Hey D. R., 2019, MNRAS, 486, 4348
Zuckerman B., 2019, ApJ, 870, 27
Zuckerman B., Bessell M. S., Song I., Kim S., 2006, ApJ, 649, L115
Zuckerman B., Klein B., Kastner J., 2019, ApJ, 887, 12

## APPENDIX A: STARS ORIGINALLY IN OUR SAMPLE BUT WERE EXCLUDED

We discarded the below stars whose measured brightness and *Gaia* colour fell within the range of our 'ZAMS + 1' sample but were later understood as possible contaminants.

(i) HD 129174 – A B9 star; its companion HD 129175 is an A6 star, 5.61 arcsec away. We suspect that *TESS* light curve for this object is a combination of both the stars, due to inability to resolve objects spaced apart <21 arcsec.

(ii) HD 105686A – It has a companion HD 105686B 3.73 arcsec away, whose *Gaia G* mag of 8.09 falls outside of our parameter range. We also suspect that *TESS* light curve for this object is a combination of both the stars, due to inability to resolve objects spaced apart <21 arcsec.

(iii) HD 35673A – It has a companion HD 35673B 3.18 arcsec away, whose *Gaia G* mag of 8.52 falls outside of our parameter range. We also suspect that *TESS* light curve for this object is a combination of both the stars, due to inability to resolve objects spaced apart <21 arcsec.

(a) HD 44770 – An F5 star; its companion HD 44769 is an A8 star, 12.31 arcsec away. The companion falls outside of our 'ZAMS + 1' sample due to its absolute magnitude, and we also suspect that *TESS* light curve for this object is a combination of both the stars, due to inability to resolve objects spaced apart <21 arcsec.

(iv) HD 77002B – Shows p modes (strong peaks in frequency range 10–16 $d^{-1}$) in the spectra, $G_{BP} - G_{RP} = -0.145$. As it is unusually blue (high surface temperature) for a $\delta$ Scuti we suspect that the *TESS* light curve is contaminated by a nearby beta cepheid ∗*b01* Car located 40.15 arcsec away.

(v) HD 6457 – A B9 star, $G_{BP} - G_{RP} = -0.59$; its companion HD 6456 is an A0 star, $G_{BP} - G_{RP} = 0.01$, 29.86 arcsec away. The light curves for both appear identical, with a $\sim$ 200 ppm strong peak at around 24 $d^{-1}$.

(vi) HD 32040 – A B9 star, displaying beating of two closely spaced frequencies, and their harmonics.